# Quasiparticle interference of heavy fermions in resonant X-ray scattering

A. Gyenis[1,†], E. H. da Silva Neto[2,3,4,5,†,‡], R. Sutarto[6], E. Schierle[7], F. He[6], E. Weschke[7], M. Kavai[8], R. E. Baumbach[9], J. D. Thompson[9], E. D. Bauer[9], Z. Fisk[10], A. Damascelli[2,3], A. Yazdani[1,*] and P. Aynajian[8,*]

[1]Joseph Henry Laboratories and Department of Physics, Princeton University, Princeton, NJ 08544, USA.
[2]Department of Physics and Astronomy, University of British Columbia, Vancouver, British Columbia V6T 1Z1, Canada.
[3]Quantum Matter Institute, University of British Columbia, Vancouver, British Columbia V6T 1Z4, Canada.
[4]Max Planck Institute for Solid State Research, Heisenbergstrasse 1, D-70569 Stuttgart, Germany.
[5]Quantum Materials Program, Canadian Institute for Advanced Research, Toronto, ON M5G 1Z8, Canada.
[6]Canadian Light Source, Saskatoon, Saskatchewan S7N 2V3, Canada.
[7]Helmholtz-Zentrum Berlin für Materialien und Energie, Albert-Einstein-Strasse 15, D-12489 Berlin, Germany.
[8]Department of Physics, Applied Physics and Astronomy, Binghamton University, Binghamton, NY 13902, USA.
[9]Los Alamos National Laboratory, Los Alamos, New Mexico 87545, USA.
[10]Department of Physics and Astronomy, University of California, Irvine, California 92697, USA.

[†] These authors contributed equally to this work.
[‡] Present address: Department of Physics, University of California, Davis, California 95616, USA.
* Corresponding author. Email: aynajian@binghamton.edu; yazdani@princeton.edu

**Abstract**

*Resonant X-ray scattering (RXS) has recently become an increasingly important tool for the study of ordering phenomena in correlated electron systems. Yet, the interpretation of the RXS experiments remains theoretically challenging due to the complexity of the RXS cross-section. Central to this debate is the recent proposal that impurity-induced Friedel oscillations, akin to quasiparticle interference signals observed with the scanning tunneling microscope (STM), can lead to scattering peaks in the RXS experiments. The possibility that quasiparticle properties can be probed in RXS measurements opens up a new avenue to study the bulk band structure of materials with the orbital and element selectivity provided by RXS. Here, we test these ideas by combining RXS and STM measurements of the heavy fermion compound $CeMIn_5$ (M = Co, Rh). Temperature and doping dependent RXS measurements at the Ce-$M_4$ edge show a broad scattering enhancement that correlates with the appearance of heavy f-electron bands in these*

*compounds. The scattering enhancement is consistent with the measured quasiparticle interference signal in the STM measurements, indicating that quasiparticle interference can be probed through the momentum distribution of RXS signals. Overall, our experiments demonstrate new opportunities for studies of correlated electronic systems using the RXS technique.*

**Introduction**

The quest to understand strongly correlated electronic states has pushed the frontiers of experimental measurements of solids to the development of new experimental techniques and methodologies. Understanding these exotic electronic states, such as those in heavy-fermions, cuprates, and pnictides, requires precise knowledge of their low energy excitations. Angle resolved photoemission spectroscopy (ARPES) and spectroscopic imaging with the scanning tunneling microscope (SI-STM) have provided a great deal of information on the nature of such low-energy states through precise measurements of their energy dispersion and interference properties (*1-4*). In the past decade, advances in resonant X-ray scattering (RXS) have provided a new tool to probe orbital-specific ordering phenomena in condensed matter systems, such as the charge order in the cuprates (*5-20*). Recently, theoretical investigations led by Abbamonte *et al.* (*21*), and subsequently by Dalla Torre *et al.* (*22*), proposed the extension of this technique to probe band structure effects, in resemblance to the quasiparticle interference signal measured using SI-STM. Remarkably, these theoretical studies demonstrate a simple and direct relation between the RXS intensity and SI-STM. Although this relation can be used to re-interpret (*22*) RXS measurements of charge order in the high-temperature superconducting cuprates (*5-20*), the equivalence between RXS and SI-STM is expected to hold more generally. Moreover, the element-specific sensitivity to the *bulk* electronic structure gives RXS a fundamental advantage over the surface sensitive STM and ARPES probes. To test these hypotheses, we carried out complementary RXS and STM studies to assess the impact and significance of quasiparticle interference to RXS experiments.

In this work, we investigate these proposed ideas on an archetypical correlated heavy fermion system and further our understanding of heavy quasiparticle formation through RXS measurements. Unconventional superconductivity and quantum criticality in *f*-electron materials develop as a consequence of heavy quasiparticle excitations emerging through the hybridization of *f*-orbitals with conduction electrons (*23-29*). Understanding these remarkable phenomena requires probing the energy-momentum structure of the emergent narrow heavy bands near the Fermi energy ($E_F$) with high precision. The CeMIn$_5$ (M = Co, Rh) family of heavy fermion compounds (*30-31*) is an ideal system for this task, as STM measurements can be carried out on these materials (*32-34*) and the energies of the Ce-$M_{4,5}$ edges (3*d* to 4*f* transition) allow RXS measurements to be performed in the currently available state-of-the-art synchrotron soft X-ray end-stations. In these Kondo lattice systems, the hybridization between the *f*-orbitals of the Ce atom and the itinerant *spd* conduction electrons leads to the formation of a narrow heavy band at the Fermi energy, below a characteristic coherence temperature $T^*$ (Fig. 1A). As a result, the heavy *f*-like band and its associated quasiparticle interference can be dramatically suppressed above $T^*$, allowing temperature to be used as a control parameter. Alternatively, isovalent substitution of the transition metal site M between Rh and Co controls the hybridization strength and consequently the large density of *f*-electron states near $E_F$ (*27*). Finally, the ground state of CeCoIn$_5$ can be modified between superconductivity and antiferromagnetism by hole doping with Cd, which enables us to study the low-energy electronic states in the vicinity of different ordered phases (*35-36*). Overall, the CeMIn$_5$ system allows the band structure to be easily tuned as a function of temperature and doping, providing an ideal test bench for the impact of band structure effects on the RXS experiment and its connection to SI-STM.

**Results**

Our earlier STM studies (*32-33*) showed that the cleaved (001)-oriented surfaces of CeCoIn$_5$ expose three different chemical terminations: surfaces A, B and C. In these previous works, we demonstrated the surface-dependent sensitivity to the heavy fermion excitations: on surface A the light quasiparticles were detected, while surface B predominantly showed the heavy quasiparticles of the hybridized band structure. Here, we carried out spectroscopic measurements in the normal state (*T* = 10 K) on surface B of CeCo(In$_{1-x}$Cd$_x$)$_5$ to study the heavy *f*- quasiparticle interference and compare it with the *4f* sensitive RXS data. Previously (*32*), the hallmark of heavy band formation was observed through the temperature-dependent large density of states in the STM spectra (Fig. 1B). In the current study, we use SI-STM to visualize its energy-momentum structure. Figures 1C-D show the real space conductance map ($x = 0.15$) at specified energies near $E_F$. The Fourier transforms of the conductance maps (Figs. 1E-F) reveal an enhancement of LDOS modulations along the [*H*, *H*] crystallographic direction. The energy momentum structure of these quasiparticles in the [*H*, *H*] direction signals the presence of rapidly dispersive bands as a function of energy (Fig. 1G). Similar results were obtained for the $x = 0.0075$ Cd doped sample (Fig. 1H). Therefore, both results reveal the formation of heavy quasiparticle bands near the chemical potential, which are independent of the Cd doping: the quasiparticle interference is unaffected by the underlying ground state at this temperature. Overall, the dispersive nature of the modulations in the STM conductance maps, with the absence of non-dispersive features, relates its origin to quasiparticle interference of heavy *f*-electrons. The quasiparticle interference originating from the heavy bands and the hybridization energy-scale are in agreement with ARPES measurements of heavy *f*-quasiparticles close to the Fermi energy (*37-38*). From the experimental perspective, however, since both RXS and STM are momentum-transfer (*Q*-space) probes, the comparison between them becomes more direct.

We now move to our RXS measurements performed on CeCo(In$_{1-x}$Cd$_x$)$_5$ samples ($x = 0$ and $x = 0.1$) and on CeRhIn$_5$. To enhance the sensitivity of our scattering measurements to the *f*-electron states, we tuned the photon energy near the Ce-$M_{4,5}$ edges, as determined by the X-ray absorption spectrum (XAS). The XAS in Fig. 2A displays peaks due to the 3*d* to 4*f* transition, with two main regions separated by $\Delta^{3d}_{SOC} \sim 17$ eV, which is the spin-orbit splitting of the $J = 3/2$ and $J = 5/2$ states of the 3*d* core-hole, and correspond to the $M_{4,5}$ edges, respectively (*39-40*). Recent dynamical mean field theory calculations (*41-42*) show the *f*-quasiparticle peak near the Fermi energy (also seen in our STM data of Fig. 1) to be entirely of $J = 5/2$ character, whereas the $J = 7/2$ *f*-band is located about 280 meV above the chemical potential. These strongly dispersing *f*-bands near the Fermi energy have been experimentally observed by ARPES measurements (*37-38*) as well as STM measurements (*33-34*). Dipole selection rules ($\Delta J = 0, \pm 1$) dictate that whereas at the $M_5$ edge both $J = 5/2$ and $J = 7/2$ unoccupied 4*f* states can be reached, only the $J = 5/2$ state can be reached at the $M_4$ edge. Therefore, we conclude that the RXS measurement at the $M_4$ edge is selectively sensitive to the narrow heavy *f*-quasiparticle peak at $E_F$, and thus allows a direct comparison to the SI-STM data of Fig. 1 in an energy window smaller than the XAS broadening. In the following, therefore, we focus on RXS measurements at the Ce-$M_4$ edge.

The RXS measurements were performed using a standard scattering geometry (Fig. 2B) with σ polarization in the (*HHL*) plane and using a photon energy $E_{ph}$ = 900.3 eV in resonance with the Ce-$M_4$ edge, unless otherwise noted. Due to the geometric limitations imposed by this small photon energy, the momentum scans are not restricted to a single value of momentum transfer *L* along the crystallographic *c*-axis (see the Supplementary Materials), similar to previous works on the cuprates (*5-11*).

Figure 3A shows momentum scans plotted as a function of in-plane momentum transfer along the two high symmetry directions [*H*, 0] and [*H*, *H*] in CeCoIn$_5$ at the Ce-$M_4$ and also at the Co-$L_2$ edge (corresponding to a 2*p* to 3*d* transition). All scans exhibit a sharp increase (decrease) of intensity for small *H* < 0.15 rlu (large *H* > 0.4 rlu) related to the geometry of the RXS experiment, where we defined rlu (reciprocal lattice unit) as 2π/*a* = 1.36 Å$^{-1}$ with *a* = 4.6 Å, the tetragonal in-plane lattice constant. Residing on top of a temperature-independent fluorescence background, the momentum scans reveal a broad scattering enhancement in the 0.2 < *H* < 0.4 rlu range, along the [*H*, *H*] direction when tuned to the Ce-$M_4$ edge, but no enhancement along the [*H*, 0] direction. Similar momentum scans, with the X-ray photon energy tuned to the Co-$L_2$ absorption edge, show the absence of the scattering enhancement along [*H*, *H*]. Furthermore, momentum scans for photon energies finely tuned around the Ce-$M_4$ resonance (Fig. 3B) demonstrate that the scattering enhancement near *H* = 0.35 rlu is resonant with the $M_4$ edge. These results indicate that the scattering enhancement observed at the $M_4$ edge along the [*H*, *H*] direction originates from heavy quasiparticles of *f*-character just above $E_F$.

Typically, resonant enhancement in scattering experiments is associated with electronic ordering. In this context it must be noted that CeCo(In$_{1-x}$Cd$_x$)$_5$ is located close to an antiferromagnetic (AFM) quantum critical point (*35-36*), with Cd doping driving the system towards the AFM ground state ($T_N$ ~ 3 K at *x* = 0.10). Hence, even though all our measurements were performed in the absence of static order, it would be conceivable that at 10 K our RXS measurements, which are energy integrated and therefore also sensitive to inelastic processes, could be picking up fluctuations of an ordered state. However, this possibility is repudiated by the insensitivity of the RXS scattering enhancement to Cd doping (Fig. 3C), demanding an alternative explanation.

To further investigate the origin of this resonant scattering enhancement, we next consider its temperature- and material-dependences. Figure 3D displays momentum scans along the [$H$, $H$] direction for several temperatures at the Ce-$M_4$ edge. The data reveal a significant temperature dependence with a rapid suppression of the scattering enhancement up to 100 K and its saturation above. Identical measurements carried out on the isostructural material CeRhIn$_5$, where heavy $f$-quasiparticles are expected to be absent from the Fermi surface at 20 K (*32*), show no temperature dependent scattering features in the same temperature window (Fig. 3E). These temperature and material dependent RXS measurements provide a direct connection to the STM measurements (Fig. 1B) and further provide an important finding. The robust presence of the heavy fermion band in CeCo(In$_{1-x}$Cd$_x$)$_5$ (for $x$ = 0.1 and 0.15) and its absence in CeRhIn$_5$ from STM and RXS indicate that the antiferromagnetic ground state, which forms at lower temperatures in both compounds, has different origins – presumably related to the itinerant or localized character of the Ce's $f$-moments. This is an important piece of information in the context of Kondo destruction and quantum criticality in heavy fermions (*43*) and deserves further investigation.

Figure 3F displays the amplitude of the scattering peak enhancement in RXS obtained at a given temperature as the difference between the area under the momentum scan at that temperature and at 200 K. Comparison to the temperature dependence of the $f$-weight from STM spectra shows a good agreement with the RXS results, indicating not only a strong correspondence between the two techniques, but also that RXS can indeed be a momentum-resolved probe of the band structure of materials beyond ordering phenomena.

Exactly how this sensitivity to the band structure occurs remains an open question in the field. To illustrate one possible scenario, in which a strongly dispersing (flat) band above the Fermi energy can give rise to a broad scattering enhancement in the RXS measurement, we follow the procedure proposed in Ref. (*21-22*), using experimentally obtained quasiparticle interference information rather than simulated data. The derived phenomenological picture relates the Fourier transform of STM conductance maps, *g(q,ω)*, to the RXS intensity (Eq. 1):

$$I_{RXS}(q, E_{ph}) = \left| \int_0^\infty \frac{g(q, \omega')}{E_{ph} - \omega' - E_h + i\Gamma_h} d\omega' \right|^2,$$

where the integral runs only over unoccupied states, while $E_h$ and $\Gamma_h$ refer to the core-hole energy and broadening, respectively. At the resonant condition (i.e. $E_{ph} = E_h$), this integral effectively relates $I_{RXS}$ to the sum of *g(q,ω)* inside an envelope of width $\Gamma_h$, which is typically ~ 200 meV.

Before applying Eq. 1 to our experimental data, we discuss the approximations and assumptions that it entails. First, we point out that choosing the proper integration boundaries is critical to the result of Eq. 1. Naturally, the limited energy range (few hundreds of meV) of an STM measurement requires us to introduce a cutoff energy for the upper bound of the integral. However, as we discussed earlier, due to dipole selection rules, the Ce-$M_4$ edge is expected to represent only the *J* = 5/2 heavy-electron states near $E_F$. Therefore, we restrict the integration only up to 20 meV, where we detect the strongly dispersing *f*-band, which is also the energy window where the tunneling spectrum exhibits a temperature dependence (Fig. 1B). Extending the integration to higher energies would suppress the strength of the signal associated with the heavy band relative to the background intensity of *g(q,ω)* and would also introduce spurious contributions from bands that may not play a role in the RXS measurement.

Secondly, it is important to note that the quasi-three-dimensional nature of the band structure of CeCoIn$_5$ is not accounted for in Eq. 1. Both STM and RXS techniques are sensitive to the $Q_x$ and $Q_y$ (or *H* and *K* in reciprocal lattice units) in-plane component of the **Q** = **k**$_i$ − **k**$_f$ momentum transfer, as well as the $Q_z$ (or *L*) out-of-plane component. Generally, quasiparticle interference maps measured by STM in a material with three-dimensional band structure can be approximated by the weighted average quasiparticle interference over $k_z$ slices (*44*), though the exact nature of this sum is not known. On the other hand, the scattering geometry in RXS measurements precisely determines the value of *L* (see the Supplementary Materials), but not the initial and final values of $k_z$. Therefore, since the sensitivity to $k_z$ may be different in the two techniques, an exact connection between RXS and STM using Eq. 1 may not hold. The calculations below should be interpreted as a qualitative description.

Figure 4 shows the calculated RXS intensity based on our STM conductance maps acquired on surface B along the [$H$, $H$] direction. We observe that the flat band indicated by the dashed area in Fig. 4A leads to a broad scattering enhancement in the $0.2 < H < 0.3$ rlu range (Fig. 4B). Quantitatively, the momentum-range of enhanced quasiparticle interference in the STM data is smaller than that seen in the RXS measurements (Fig. 4C). This difference could be the result of the quasi-three-dimensionality of the CeCoIn$_5$ band structure as discussed above. Regardless, since the integration is taken over fastly dispersing (heavy) bands, the RXS scattering signal predicted using Eq. 1 is expected to be broad in momentum space – similar to the RXS data in Fig. 3.

**Discussion**

Here we showed the results of a complementary SI-STM and RXS study that probes the significance of band structure effects in RXS scattering measurements. The temperature-, material- and photon-energy-dependences of the RXS data clearly indicate its sensitivity to the formation of the Kondo lattice in the CeMIn$_5$ system. Our model calculations based on the experimental data show that the strongly dispersing $f$-bands can give rise to an enhancement of the RXS in a similar momentum range. These observations suggest that bulk quasiparticle interference, as proposed in Refs. (*21-22*), are responsible for the RXS signal in the present measurements. We should emphasize that whereas quasiparticle interference in STM conductance maps arises from native defects, atomic step edges and impurities, in RXS measurements it may have an additional contribution. When an X-ray photon excites a core electron to the valence band, it creates a localized core hole potential, which can act as a scattering center. Therefore, this perturbation can also lead to the scattering and interference of the itinerant quasiparticles (*45*).

It must be also noted that orbital degrees of freedom and their structure factors play a key role in RXS and SI-STM measurements, as in the case of the $d$-wave form factor of the charge order in the cuprates (*12-13*). To obtain the relation between RXS and SI-STM expressed in Eq. 1, Abbamonte *et al*. (*21*) purposely disregarded the atomic orbital components of the electronic wave function. This is a particularly inappropriate approximation for *f*-electron systems where the atomic multiplet structure is usually well-defined in core-hole spectroscopies. This opens the possibility, for example, that the broad scattering enhancement in the RXS is related to how the polarized light couples to particular orbitals (*39*), which might only become available to the X-ray scattering process after band hybridization below $T^*$. In our experiments, this scenario seems unlikely, given that all our measurements were done in σ-scattering geometry (i.e. light polarization constantly in the *a-b* plane of the sample throughout the momentum scans, as Fig. 2B shows). This suggests that the momentum structure observed in our RXS data is likely related to the band structure of $CeCoIn_5$ – as supported by the temperature dependent measurements and their similarity to the SI-STM signal. Nevertheless, given the complexity of the RXS cross-section, a comprehensive treatment of the scattering process, which takes into account the atomic multiplets in *f*-electron systems, would provide a more natural explanation to our results.

Our experiments, which demonstrate the relevance of quasiparticle interference in RXS, may also be relevant to the cuprates. In the past few years, a universal charge order instability emerged as the most exciting progress in the study of high-temperature superconductivity in cuprates (*5-20*). The ubiquitous nature of this electronic phenomenon in the bulk of hole- and electron-doped cuprates came from RXS experiments on $(Y,Nd)Ba_2Cu_3O_y$, Bi-2201, Bi-2212, $HgBa_2CuO_{4+\delta}$, and $Nd_{2-x}Ce_xCuO_4$ (*6-11*). These measurements reveal an incommensurate scattering peak, with correlation lengths ranging from 20 to 75 Å depending on the material and doping. Making a parallel between our RXS experiment on $CeCoIn_5$ to those on the cuprates, suggests that bulk quasiparticle interference features might be present in the RXS signal of the latter, perhaps even in coexistence with the charge order peak. At this point, only further experiments can clarify the impact of quasiparticle interference to the RXS measurements in the cuprates.

Most importantly, our experiments demonstrate that RXS can be a powerful momentum- and energy-resolved probe of the bulk band structure of materials, even in the absence of any ordering phenomena. These results not only pave the way to future RXS experiments on *f*- electron materials, but also support the complementary relationship between RXS and STM measurements.

**Methods: sample growth, STM and RXS measurement technique**

The single-crystal samples used for the measurements were grown from excess indium at Los Alamos National Laboratory. Small, flat crystals were oriented along the crystallographic axes and cut into sizes suitable for STM and RXS measurements (0.5-2 mm x 0.5-2 mm x 0.2 mm). The samples used for the STM measurements were cleaved perpendicular to the *c*-axis in ultra-high vacuum at room temperature and immediately inserted into our home-built variable temperature STM. Differential conductance measurements were performed using standard lock-in techniques, with bias applied to the sample.

The reported RXS experiments were performed at the REIXS beamline of the Canadian Light Source using a 4-circle diffractometer in the temperature range of 22 to 200 K (*46*) and at the UE46-PGM1 beam line of the Helmholtz-Zentrum Berlin at BESSY-II with a 2-circle diffractometer between 10 K and 200 K. Reciprocal-space scans were acquired by rocking the sample angle ($\theta$) at fixed detector position ($\theta_{det}$=170°). The samples were pre-oriented using Laue diffraction.

**Acknowledgments**

We thank Robert Green and George Sawatzky for stimulating discussions about the resonant X-ray process and HZB for the allocation of synchrotron radiation beamtime. The work at Princeton was primarily supported by a grant from DOE-BES and Gordon and Betty Moore Foundation as part of EPiQS initiative (GBMF4530). The instrumentation and infrastructure at the Princeton Nanoscale Microscopy Laboratory used for this work were also supported by grants from NSF-DMR1104612, the NSF-MRSEC program through Princeton Center for Complex Materials (DMR-1420541), the Linda and Eric Schmidt Transformative Fund, and the W. M. Keck Foundation. The work at Binghamton University was supported by P.A.'s startup funds. The work at Los Alamos was performed under the auspices of the U. S. Department of Energy, Office of Basic Energy Sciences, Division of Materials Sciences and Engineering. Part of the research described in this paper was performed at the Canadian Light Source, which is supported by the Canada Foundation for Innovation, Natural Sciences and Engineering Research Council of Canada, the University of Saskatchewan, the Government of Saskatchewan, Western Economic Diversification Canada, the National Research Council Canada, and the Canadian Institutes of Health Research. The work at UBC was supported by the Max Planck-University of British Columbia Centre for Quantum Materials; the Killam, Alfred P. Sloan, and NSERC's Steacie Memorial Fellowships; the Alexander von Humboldt Fellowship; the Canada Research Chairs Program; and the Natural Sciences and Engineering Research Council of Canada (NSERC), Canada Foundation for Innovation (CFI), and the Canadian Institute for Advanced Research's (CIFAR) Quantum Materials program. E.H.d.S.N. acknowledges support from the CIFAR Global Academy.

**Author contributions**

A. G., P. A. and E. H. d. S. N. performed the STM measurements. E. H. d. S. N., A. G., P. A, R. S., E. S., F.H. and E. W. performed the RXS measurements. A. G., E. H. d. S. N., M. K. and P. A. analyzed the data. R. E. B., J. D. T., Z. F. and E. D. B. synthesized and characterized the materials. P. A., E. H. d. S. N., A.G., and A. Y. wrote the manuscript. All authors commented on the manuscript.

The authors declare that they have no competing interest. All data needed to evaluate the conclusions in the paper are present in the paper and/or the Supplementary Materials. Additional data related to this paper may be requested from the authors.

**Figure Captions**

**Fig. 1. STM studies on heavy fermion $CeCoIn_5$.** (**A**) Illustration of heavy fermion band formation as a result of hybridization below the $T^*$ coherence temperature of the Kondo lattice. (**B**) Temperature dependence of the averaged tunneling spectra on surface B of pure $CeCoIn_5$ and $CeRhIn_5$ (dashed line). Data is from (*32*). (**C** and **D**) Real space conductance map near the Fermi energy on surface B of $CeCo(In_{1-x}Cd_x)_5$ at $x = 0.15$ doping level, which show clear heavy-quasiparticle interference waves ($V_{bias} = -100$ mV, $I_{setpoint} = 1.6$ nA). (**E** and **F**) Fourier transforms of the real space conductance maps at the corresponding energies, which display dispersing peaks in the [$H$, $H$] direction. Red dot indicates the (0.4, 0.4) point in the reciprocal space. (**G** and **H**) Energy-momentum cuts of the Fourier transforms in the [$H$, $H$] direction (dashed line on E) for $x = 0.15$ and $x = 0.0075$. The heavy fermion band formation and quasiparticle interference is unaffected by the Cd doping.

**Fig. 2. Resonant transition of the heavy 4f state.** (**A**) Linearly polarized XAS spectra (measured through the total electron yield - TEY) on $CeCo(In_{1-x}Cd_x)_5$ at $x = 0$ (blue) and $x = 0.1$ (green) doping

level at 10 K, which demonstrates the transition between the 3d and 4f states. The curves are shifted vertically for clarity. (**B**) RXS scattering geometry with the scattering plane lying at 45° relative to the *a* and *b* axes.

**Fig. 3. RXS measurements on CeCoIn$_5$.** (**A**) On-resonance θ scans showing the RXS diffraction signal on pure CeCoIn$_5$ at Ce-$M_4$ edge in both the [*H*, *H*] and [*H*, 0] directions, and as a comparison at the Co-$L_2$ edge in the [*H*, *H*] direction at *T* = 22 K. The curves are shifted vertically for clarity. (**B**) Energy-dependence of the RXS spectrum on and off the Ce-$M_4$ edge. Inset shows the XAS around the $M_4$ edge with arrows indicating the energies where RXS spectra were measured. (**C**) Comparison of the RXS cross section at the $M_4$ edge on CeCo(In$_{1-x}$Cd$_x$)$_5$ at *x* = 0 (red) and *x* = 0.1 (blue) doping level at *T* = 10 K. The RXS signal exhibits the same shape independently from the doping level. (**D**) Detailed temperature evolution of the RXS peak on pure CeCoIn$_5$, which reveals strong temperature-dependence. (**E**) RXS scans measured on CeRhIn$_5$ at *T* = 20 K and *T* = 100 K. The cross sections display no significant temperature-dependence. (**F**) Comparison the RXS peak height with the hybridization peak height measured by STM (from Fig. 1B) as a function of temperature. Both signals exhibit a sharp up-rise at the *T\** Kondo temperature suggesting the common origin of the features.

**Fig. 4. Quasiparticle interference and RXS.** (**A**) Energy-momentum structure of the quasiparticle band in the [*H*, *H*] direction on surface B with white dashed lines indicating the boundary of the integral (Eq. 1) used to calculate the RXS intensity. (**B**) Calculated RXS intensity which exhibits a broad kink starting around *H* = 0.2 rlu. The momentum range, where the broad peak appears, is indicated by green dashed lines both on panel (A) and (B). (**C**) Experimentally measured RXS intensity on CeCoIn$_5$ at *T* = 22 K, which shows a broad peak around 0.2 < *H* < 0.4 rlu. The corresponding momentum range is indicated by dashed lines.

Figure 1

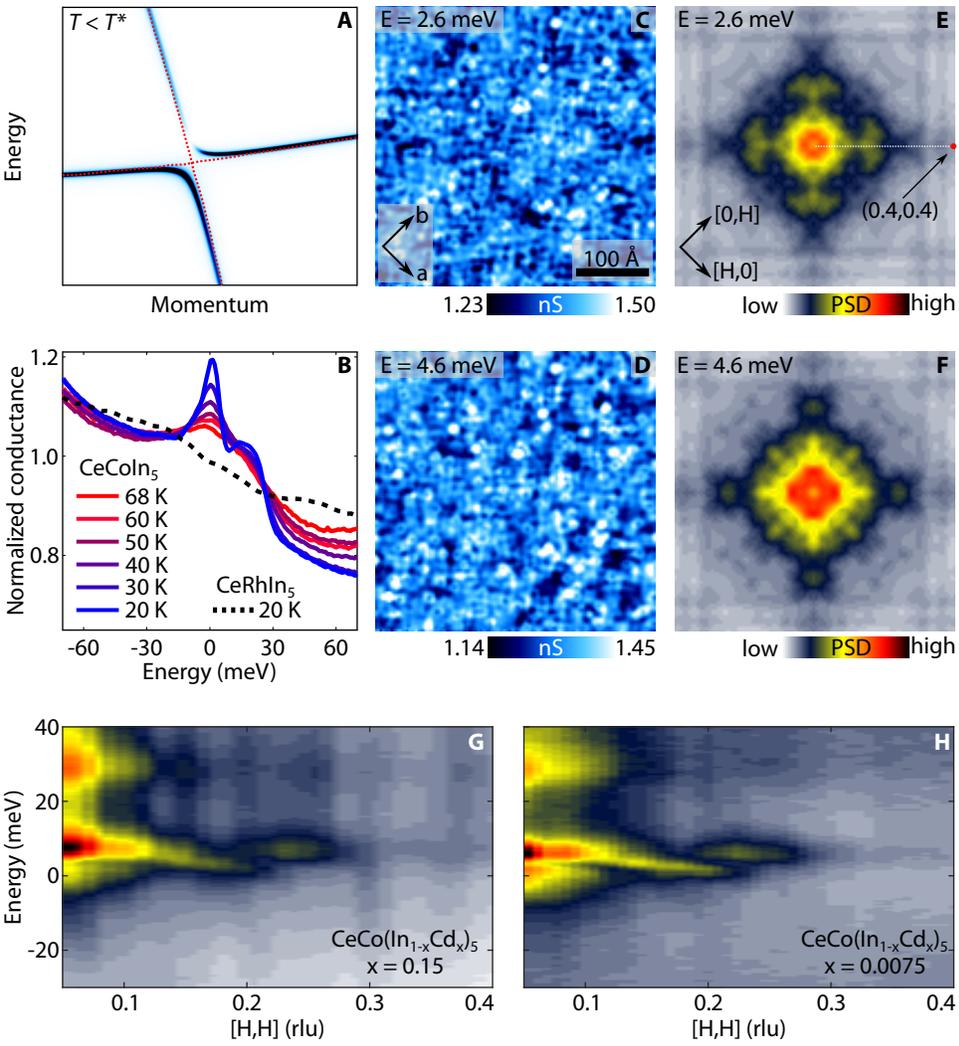

Figure 2

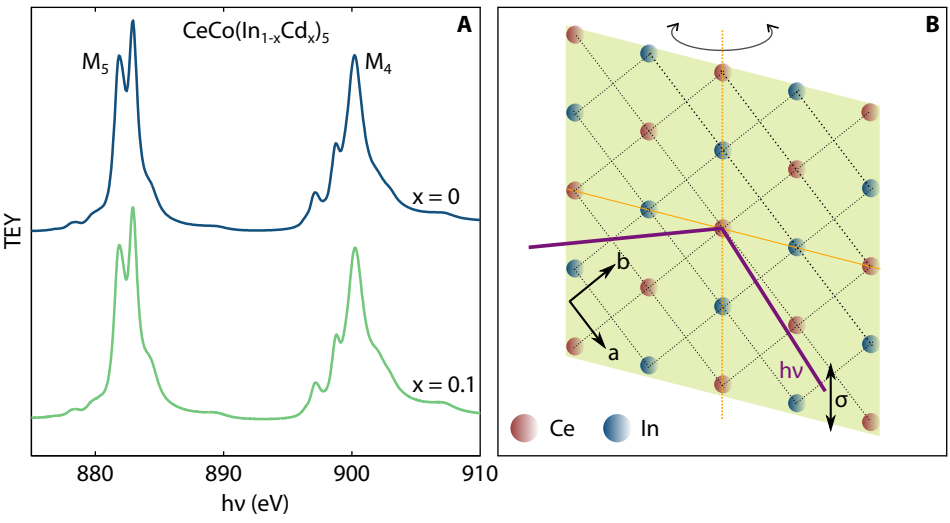

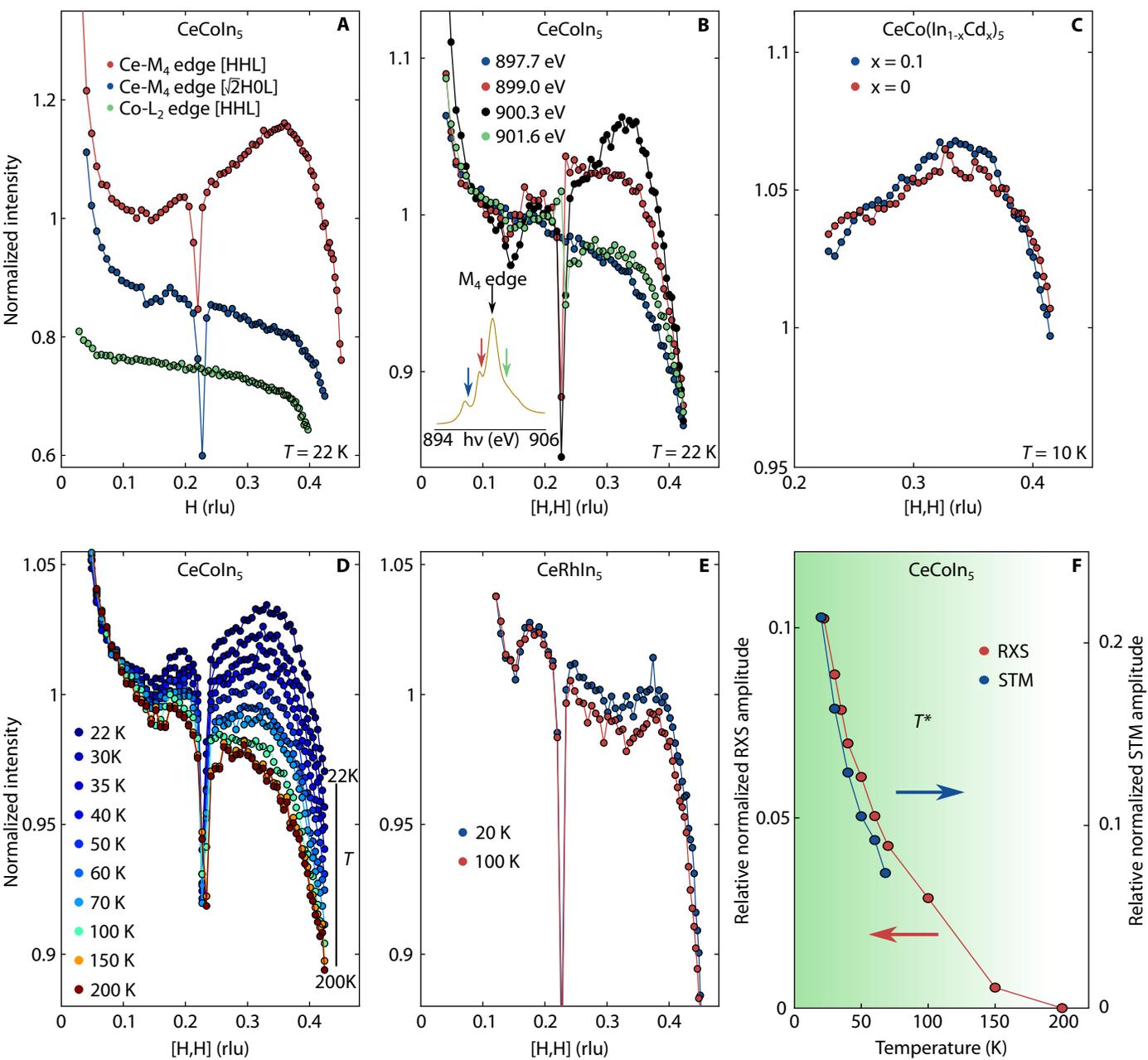

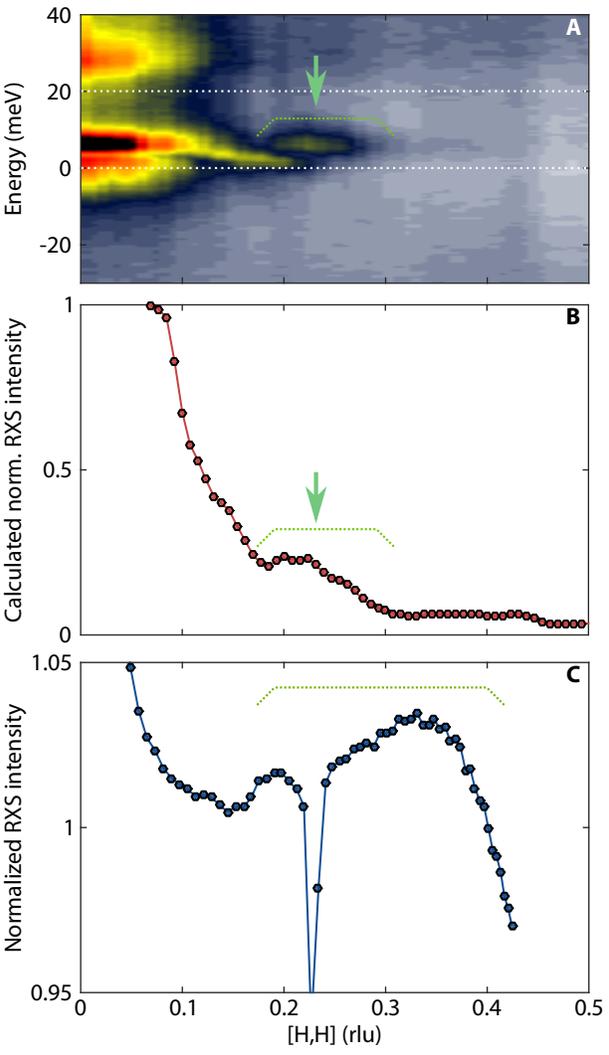

# Supplementary Materials for

Quasiparticle interference of heavy fermions in resonant X-ray scattering

A. Gyenis,[†] E. H. da Silva Neto,[†] R. Sutarto, E. Schierle, F. He, E. Weschke, M. Kavai,

R. E. Baumbach, J. D. Thompson, E. D. Bauer, Z. Fisk, A. Damascelli, A. Yazdani,[*] and P. Aynajian[*]

[†] These authors contributed equally to this work.

* Corresponding authors. Email: aynajian@binghamton.edu; yazdani@princeton.edu

STM measurements and data analysis

The reported STM conductance maps were acquired on a 475 Å x 475 Å ($x = 0.15$) and on a 400 Å x 400 Å ($x = 0.0075$) large square area (Figs. S1A-B), by applying lock-in oscillation of 0.5 meV at parking bias of -100 meV and setpoint current of 1.6 nA ($x = 0.15$) and 1 nA ($x = 0.0075$). The features were reproduced at different tunneling junction impedances. The bright spots on the topographic images correspond to the Cd dopants. The spatially averaged $dI/dV$ spectra on both samples show the double-peak feature (Fig. S1C) demonstrating the heavy band formation at $T = 10$ K independently of the underlying ground state in $CeCo(In_{1-x}Cd_x)_5$.

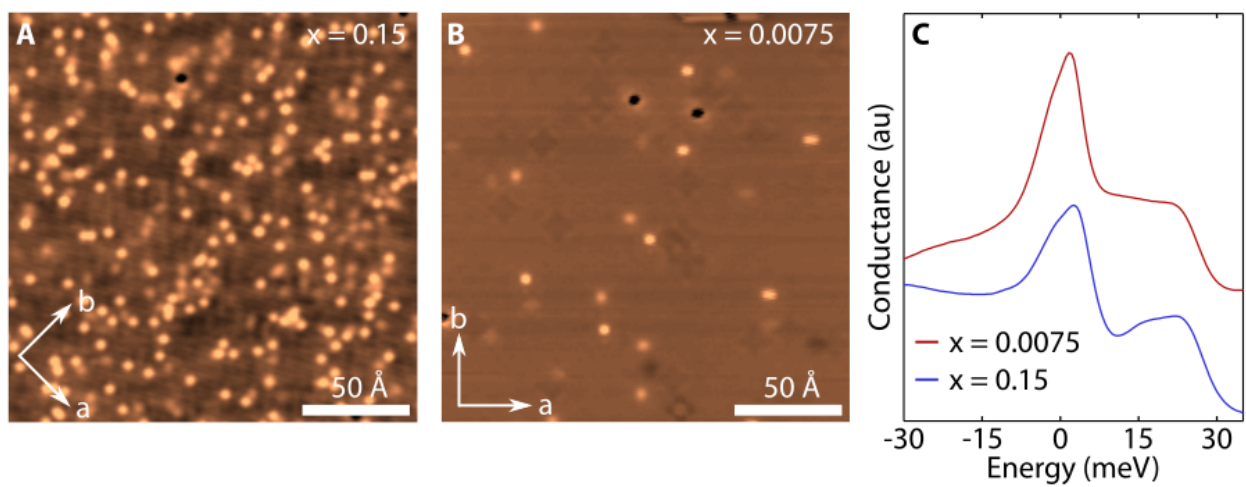

**Fig. S1. STM measurements on $CeCo(In_{1-x}Cd_x)_5$.** Topographic image of (**A**) $x = 0.15$ and (**B**) $x = 0.0075$ Cd-doped samples. (**C**) Spatially averaged tunneling conductance on surface B on the two samples.

In the real space conductance maps, the conductance values on the Cd impurities observable on the topographic image were replaced by the average conductance value at the given energy to suppress the high / low tunneling value into a bound state of the Cd dopant, which is independent from the quasiparticle interference. The real space maps were normalized by their mean prior to taking the Fourier transformation, which allows us to directly compare the modulation strength at different energies. Finally, the Fourier transform of the maps were symmetrized based on the symmetry of the underlying lattice (mirror symmetry along the *a* and *b* axis and also 90 degrees rotation along the *c* axis) to eliminate the effects of the random shape of the STM tip (Fig. S2). We note that the reported features are apparent without any of the manipulation and the described procedure serves only the purpose of enhancing the signal to noise ratio.

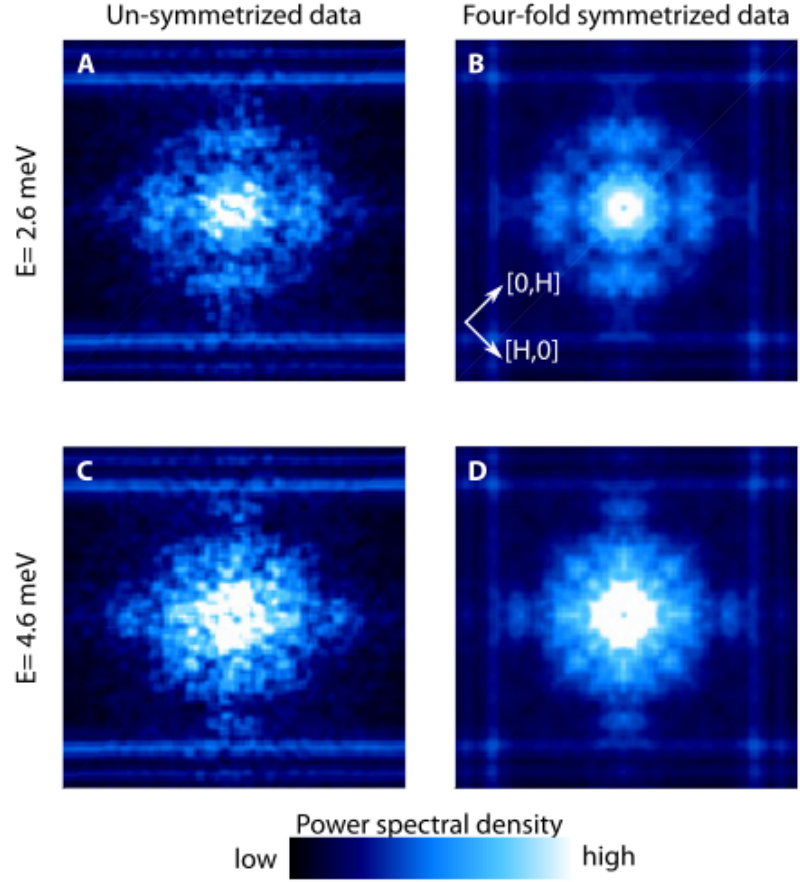

**Fig. S2. Symmetrization of the conductance map.** (**A**) and (**C**) Un-symmetrized Fourier transform of the real space conductance maps shown in Figs. 1C-D. (**B**) and (**D**) Four-fold symmetrized conductance map on a single colorscale.

RXS comparison on Cd doped sample

As we discussed it in the main text, we observed the same RXS peak enhancement in case of pure and 10% Cd doped sample (Fig. 3C). Here, we show additional temperature dependence of the RXS peaks, which exhibit remarkably similar behavior. As Fig. S3 demonstrates, the RXS peak is absent above the hybridization temperature $T^*$ in both samples.

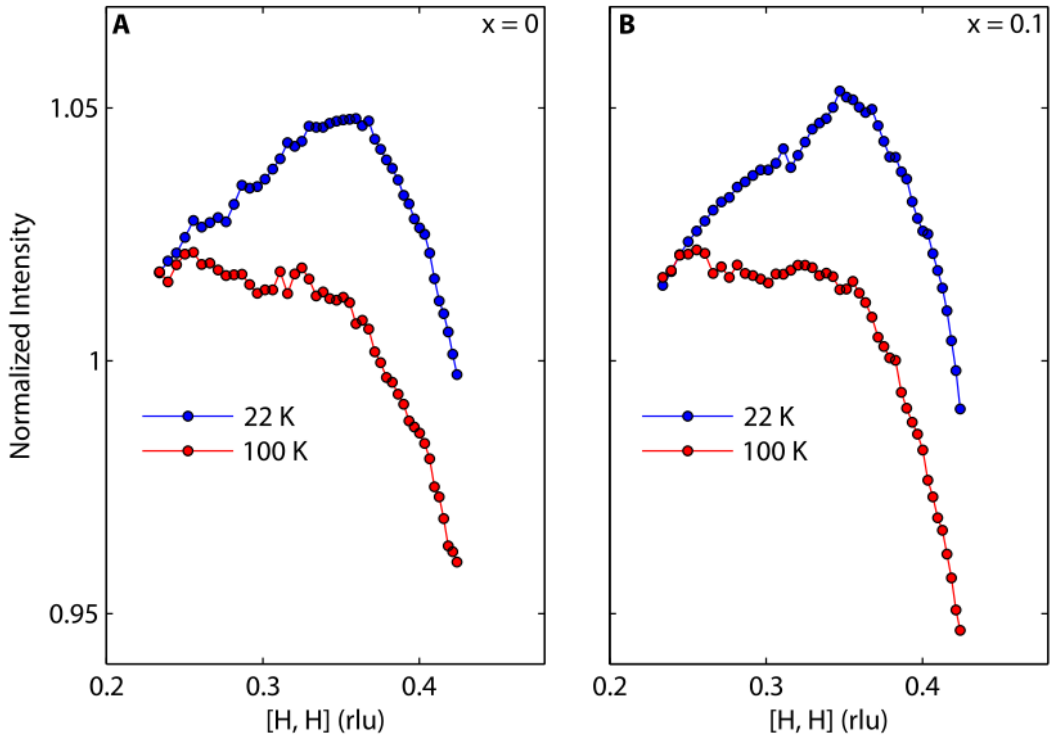

**Fig. S3. Comparison of the RXS cross section on CeCo(In$_{1-x}$Cd$_x$)$_5$.** Temperature dependence of the RXS signal on (**A**) pure and (**B**) $x$ = 0.1 Cd doped samples at $T$ = 22 K and 100 K.

L dependence of the RXS enhancement

Since the momentum scans are not restricted to a single $L$ value, our RXS measurement probes the scattering signal on a cut [$H(\theta_{sample}, \theta_{det})$, $H(\theta_{sample}, \theta_{det})$, $L(\theta_{sample}, \theta_{det})$] through reciprocal space determined by the sample ($\theta_{sample}$) and detector angles ($\theta_{det}$). Therefore, it is more complicated than the surface investigated by STM. Figure S4 displays the $L$ dependence of the scattering peak on the top axis of the plot, which shows that the broad peak exists on a wide range of $L$. More importantly, since CeCoIn$_5$ has a three-dimensional band-structure, the two techniques might measure different $k_z$ components of the band structure and a truly direct comparison of the two results is not possible, as it is discussed in details in the main text.

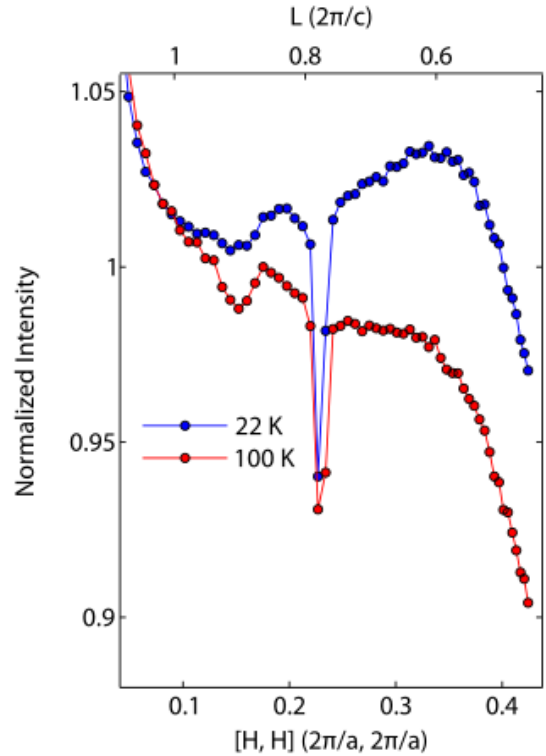

**Fig. S4. L dependence of the RXS measurement.** The observed RXS peaks at $T$ = 22 K and 100 K with their $L$ dependence indicated on the top axis.

Anomalous drop in the RXS signal

As Fig. 3 in the main text shows, the measured RXS scattering amplitude exhibits a sharp drop near $H$ = 0.22 rlu along the [$H$, $H$] direction. Although the origin of this feature is unknown, its relative sharpness, temperature- and doping-independence suggest that it is related to some kind of destructive interference in the scattering process.